\documentclass[10pt,conference, letterpaper]{IEEEtran}

\usepackage[T1]{fontenc}
\usepackage{graphicx}
\usepackage{comment}
\usepackage{cite}
\usepackage{amsmath,amssymb,amsfonts}
\usepackage{algorithmic}
\usepackage{graphicx}
\usepackage{textcomp}
\usepackage{colortbl}
\usepackage[table]{xcolor}
\def\BibTeX{{\rm B\kern-.05em{\sc i\kern-.025em b}\kern-.08em
		T\kern-.1667em\lower.7ex\hbox{E}\kern-.125emX}}

\usepackage[a-2b]{pdfx}

\usepackage{tabularx}

\usepackage[utf8]{inputenc}
\usepackage[T1]{fontenc}

\usepackage[abbreviations]{foreign}

\usepackage{graphicx}
\usepackage[margin=3pt,skip=3pt,belowskip=-10pt,font={small,stretch=0.9},labelfont=bf]{caption}

\usepackage[compact]{titlesec}
\usepackage{microtype}
\usepackage{setspace}
\usepackage{listings}
\usepackage{multirow}
\usepackage{scrextend}

\usepackage{bigdelim} 

\usepackage{tikz}
\usepackage{xspace}

\usetikzlibrary{shapes, arrows, positioning, fit, backgrounds, calc}

\usepackage{subcaption}
\usepackage{textcomp}
\usepackage{booktabs}
\usepackage{url}
\usepackage{pifont}
\usepackage{hyperref}
\usepackage{lipsum}

\usepackage[inline]{enumitem}

\usepackage[capitalise]{cleveref}

\usepackage{tikz}

\definecolor{lightcolor}{rgb}{0,0.5,1}
\usepackage[switch]{lineno} 

\usepackage{fontawesome5}
\newboolean{showcomments}
\setboolean{showcomments}{false}
\ifthenelse{\boolean{showcomments}}
{ \newcommand{\mynote}[3]{
		\fbox{\bfseries\sffamily\scriptsize#1}
		{\small$\blacktriangleright$\textsf{\emph{\color{#3}{#2}}}$\blacktriangleleft$}}}
{ \newcommand{\mynote}[3]{}}

\newcommand{\vs}[1]{\mynote{Valerio}{#1}{blue}}
\newcommand{\py}[1]{\mynote{Peterson}{#1}{red}}
\newcommand{\er}[1]{\mynote{Etienne}{#1}{purple}}
\newcommand{\emi}[1]{\mynote{Emi}{#1}{olive}}

\newcommand{\aaa}[1]{\mynote{Andrei}{#1}{magenta}}

\definecolor{darkgreen}{rgb}{0.3,0.5,0.3}
\definecolor{darkblue}{rgb}{0.3,0.3,0.5}
\definecolor{darkred}{rgb}{0.5,0.3,0.3}

\newcommand{\nopbsnotee}{\texttt{\small\textsf{noPBS-noTEE}}\xspace}
\newcommand{\nopbstee}{\texttt{\small\textsf{noPBS-TEE}}\xspace}
\newcommand{\pbsnotee}{\texttt{\small\textsf{PBS-noTEE}}\xspace}
\newcommand{\pbstee}{\texttt{\small\textsf{PBS-TEE}}\xspace}

\newcommand{\pbsteerb}{\texttt{\small\textsf{PBS-TEE-rb}}\xspace}

\newcommand{\opgethbuilder}{\texttt{\small\textsf{op-geth}}\xspace}
\newcommand{\oprbuilder}{\texttt{\small\textsf{op-rbuilder}}\xspace}

\lstdefinestyle{CStyle}{
	backgroundcolor=\color{backgroundColour},   
	commentstyle=\color{mGreen},
	keywordstyle=\color{magenta},
	numberstyle=\tiny\color{mGray},
	stringstyle=\color{mPurple},
	basicstyle=\footnotesize,
	breakatwhitespace=false,         
	breaklines=true,                 
	captionpos=b,                    
	keepspaces=true,                 
	numbers=left,                    
	numbersep=5pt,                  
	showspaces=false,                
	showstringspaces=false,
	showtabs=false,                  
	tabsize=2,
	language=C
}

\newcounter{numobserv} 
\definecolor{beaublue}{rgb}{0.88, 0.93, 0.93}
\usepackage{tcolorbox}
\colorlet{shadecolor}{beaublue}
\tcbset{
	colback=beaublue,
	boxrule=0.5pt,
	boxsep=0pt,
	left=1pt,
	right=1pt,
	top=1pt,
	bottom=1pt,
	sharp corners,
	colframe=white,
}
\newcommand{\observ}[1]{
	\addtocounter{numobserv}{1}
	\begin{tcolorbox}	
		\textit{\textbf{Take-away\,\thenumobserv\,:} #1 }	
	\end{tcolorbox}
}

\definecolor{mygreen}{RGB}{0,100,0}
\definecolor{myred}{RGB}{128,0,32}

\begin{document}
\title{Empirical Analysis of TEE-Enabled Rollup Designs}
%

\makeatletter
\newcommand{\linebreakand}{%
\end{@IEEEauthorhalign}
\hfill\mbox{}\par
\mbox{}\hfill\begin{@IEEEauthorhalign}
}
\makeatother

\let\oldIEEEauthorblockN\IEEEauthorblockN
\renewcommand{\IEEEauthorblockN}[1]{%
  \oldIEEEauthorblockN{\small #1}%
}

\let\oldIEEEauthorblockA\IEEEauthorblockA
\renewcommand{\IEEEauthorblockA}[1]{%
  \oldIEEEauthorblockA{\small #1}%
}

\author{
\IEEEauthorblockN{Claudiu-Nicu B\u{a}rbieru}
\IEEEauthorblockA{
\textit{Alexandru Ioan Cuza University}\\
Ia\c{s}i, Romania\\
claudiu.barbieru@gmail.com
}
\and
\IEEEauthorblockN{Corentin Libert}
\IEEEauthorblockA{
\textit{UCLouvain}\\
Louvain-la-Neuve, Belgium\\
corentin.libert@uclouvain.be 
}
\and 
\IEEEauthorblockN{Peterson Yuhala}
\IEEEauthorblockA{
\textit{The University of British Columbia}\\
Vancouver, BC, Canada\\
pyuhala@cs.ubc.ca
}
\and
\IEEEauthorblockN{Andrei Arusoaie}
\IEEEauthorblockA{
\textit{Alexandru Ioan Cuza University}\\
Ia\c{s}i, Romania\\
andrei.arusoaie@uaic.ro
}

\linebreakand

\IEEEauthorblockN{Matthieu Bettinger}
\IEEEauthorblockA{
\textit{UCLouvain}\\
Louvain-la-Neuve, Belgium\\
matthieu.bettinger@uclouvain.be 
}
\and
\IEEEauthorblockN{Oana-Otilia Captarencu}
\IEEEauthorblockA{
\textit{Alexandru Ioan Cuza University}\\
Ia\c{s}i, Romania\\
oana.captarencu@info.uaic.ro
}
\and
\IEEEauthorblockN{Pascal Felber}
\IEEEauthorblockA{
\textit{University of Neuch\^{a}tel}\\
Neuch\^{a}tel, Switzerland\\
pascal.felber@unine.ch
}
\and
\IEEEauthorblockN{Emanuel Onica}
\IEEEauthorblockA{
\textit{Alexandru Ioan Cuza University}\\
Ia\c{s}i, Romania\\
emanuel.onica@uaic.ro
}

\linebreakand

\IEEEauthorblockN{Etienne Rivi\`{e}re}
\IEEEauthorblockA{
\textit{UCLouvain}\\
Louvain-la-Neuve, Belgium\\
etienne.riviere@uclouvain.be 
}
\and
\IEEEauthorblockN{Valerio Schiavoni}
\IEEEauthorblockA{
\textit{University of Neuch\^{a}tel}\\
Neuch\^{a}tel, Switzerland\\
valerio.schiavoni@unine.ch
}

}

\maketitle              
\vspace{-0.25cm}
\begin{abstract}
Layer 2 (L2) rollups have become the dominant scaling solution in Ethereum. 
At the core of these systems lies the rollup sequencer, a component that orders, executes, and submits L2 transaction blocks to the main Ethereum blockchain.
The centralization of this sequencer introduces security risks, including censorship and malicious manipulation of transactions to extract maximum extractable value (MEV).
Trusted execution environments (TEEs) provide strong hardware-enforced confidentiality and integrity guarantees for code and data.
TEEs are strong candidates for strengthening security and ensuring verifiable neutrality in rollup sequencer operations, but multiple TEE-based designs are possible, offering different trade-offs among modularity, performance, and security guarantees.
Among them, the separation of proposer and builder logic (PBS) into independent services is a strong contender against the monolithic approach that bundles all sequencer logic.
Despite several proposed theoretical frameworks, there is a lack of experimental evaluation of the practicality and security trade-offs of integrating TEEs into rollup sequencers. 
This work addresses this gap through a comprehensive empirical study of TEE-enabled rollup designs under both PBS and monolithic sequencer models. 
Our experimental evaluations show that the impact of TEEs on rollup performance is small and that PBS does not significantly amplify it. 
We also show that PBS can lead to significant performance improvements by enabling more specialized, efficient builder logic implementations.
\end{abstract}


\section{Introduction}
\label{sec:intro}


Ethereum~\cite{wood2014ethereum} is a widely popular blockchain, the first to support rich applications beyond simple financial transactions. 
Its core technology, the Ethereum Virtual Machine (EVM), is now used by more than 40 other blockchains~\cite{l2beat}. 
The Ethereum Mainnet, also known as the Layer-1 (L1) Ethereum network, has faced scalability limitations in recent years, primarily due to rising transaction processing loads, which affect throughput and finality~\cite{scaling}. 
To address these limitations, Layer-2 (L2) chains, and in particular \emph{rollups}~\cite{thibault22}, have emerged as the primary scaling solution for Ethereum. 
Rollups seek to perform as much computation as possible off-chain, aggregating blocks of transactions into batches that are then submitted to L1 to inherit its decentralized-consensus guarantees.

A central component of a rollup is the \emph{sequencer}~\cite{motepalli2023sok}, which orders and executes transactions and assembles them into L2 blocks. 
These blocks are compressed, batched, and posted to the underlying L1 blockchain by a \textit{batcher} component, which is sometimes integrated with the sequencer.
In most existing rollups, the sequencer is run by a single administrative entity, \eg The Optimism Foundation~\cite{optimism_protocol_overview} or Arbitrum~\cite{arbitrum}.
This centralization introduces risks of censorship~\cite{10.1145/3618257.3624824}.
Blockchains are increasingly used beyond decentralized finance, including in e-government~\cite{GovDApps} and supply chain management~\cite{supplychain}. 
Transaction censorship can therefore severely affect the safety of such applications: censored orders can disrupt financial exchanges, dropped votes can corrupt electronic polling, and missing shipment registrations can alter supply pipelines. 
Beyond directly blocking transactions to or from specific parties, sequencers may also act for financial gain by launching frontrunning attacks~\cite{DBLP:conf/uss/TorresCS21} or manipulating transaction ordering to extract profit, known as \emph{maximum extractable value} (MEV)~\cite{alipanahloo2024maximum}.


\emph{Proposer–builder separation} (PBS)~\cite{ethereum-pbs, 10.1145/3618257.3624824} emerged as a solution to mitigate the security and scalability challenges in L1 by separating block construction (performed by builders) from block validation (performed by proposers). 
MEV-Boost~\cite{mev-boost}, Ethereum L1's dominant PBS implementation, enables a competitive builder market that allows builders to capture MEV transparently.
Builders compete to assemble high-quality blocks that maximize MEV and priority fees legitimately, retaining the surplus after bidding to proposers, while proposers earn revenue from builder bids in addition to staking rewards. 

Looking ahead, an L2 builder market is envisioned as a possible future direction that could reduce the censorship power currently concentrated in sequencers. 
However, today’s rollup deployments remain far from this model: they do not expose an L1-style PBS mechanism with an open, competitive builder market. 
What has been adopted so far, following the Rollup Boost proposal~\cite{intro-rollup-boost}, is a limited 1-to-1 sequencer--builder separation in several rollups~\cite{unichain, opmainnet-flashblocks, base-flashblocks}. 
In this configuration, the sequencer remains logically centralized, while a single builder is used de facto to support performance-oriented functionality, such as precomputing partial blocks (flashblocks) and leveraging faster modular implementations~\cite{op-rbuilder}. 
Thus, current L2 deployments should be understood as PBS-inspired architectures rather than as fully fledged PBS implementations with an open or competitive builder market.

While PBS may, in the future, support fairer MEV distribution, stronger decentralization, and improved performance in L2 systems, in the current 1-to-1 setting, the relevant censorship risks are essentially transferred from the sequencer to the builder side~\cite{10.1145/3618257.3624824}, thereby preserving similar safety concerns.

Trusted execution environments (TEEs) are hardware-based mechanisms ensuring code and data confidentiality and integrity during execution. 
Notable examples include Intel Software Guard Extensions (SGX)~\cite{vcostan}, Intel Trust Domain Extensions (TDX)~\cite{cheng24}, and AMD Secure Encrypted Virtualization (SEV)~\cite{sev-snp}. 
These TEEs provide attestation capabilities~\cite{menetrey2022attestation}, allowing remote parties to verify the executing code and authenticate the underlying hardware.
This enables verifying if a block builder's code complies with stated ordering rules, thereby averting censorship by malicious nodes.

Prior work explored TEE-enforced transaction ordering in L1~\cite{li25} or TEE-enabled sequencer networks in L2~\cite{DBLP:journals/tc/WenFLNZF25}.
However, we lack a thorough analysis of the performance implications of integrating VM-level TEEs into PBS-based L2 designs that have gained traction in production~\cite{unichain, opmainnet-flashblocks, base-flashblocks}. 
We present the first experimental study of such TEE-enabled rollup designs in both PBS and non-PBS settings.
Our evaluations with VM-based TEEs show that their performance impact is small for lightweight ERC-20~\cite{erc-20} based workloads, the most implemented Ethereum token for fungible assets, with more than 1.7M tracked instances~\cite{tokenstotal} and close to 2 billion trades executed between 2015 and 2024~\cite{somin2025cryptoasset}. 

Overall, our key findings include the following: \er{revise these based on the final evaluations.}

\aaa{proposed something below}
\begin{enumerate}
\item VM-based TEEs have a negligible impact on transaction time to inclusion in both PBS and non-PBS designs;
\item PBS enables more specialized, efficient block builders with significant performance gains over a monolithic sequencer;
\item TEEs increase CPU usage by about $17\%$ and $25\%$ for non-PBS and PBS setups, mainly due to hardware-based cryptographic operations in TEEs;
\item In PBS settings, communicating with an external builder increases CPU usage for block construction;
\item RAM utilization is largely dominated by the block builder implementation rather than PBS and TEE usage.
\end{enumerate}


\newcommand{\mysection}{Section~}

\textbf{Roadmap.} We structure our paper as follows.
\mysection\ref{sec:background} discusses background concepts like L2 rollups, PBS and TEEs.
\mysection\ref{sec:arch} explores the rollup designs our paper focuses on, while \mysection\ref{sec:eval} presents our empirical evaluation of these designs and provides useful takeaways.
\mysection\ref{sec:rw} discusses related work, and \mysection\ref{sec:conclusion} concludes our paper.

\section{Background}
\label{sec:background}

In what follows, we provide details on the major rollup categories, elaborate on the centralized state of PBS in L2, and summarize the different types of TEEs.

\subsection{L2 rollups}
Rollups are categorized into \emph{ZK-rollups} and \emph{optimistic rollups} based on the transaction validity verification mechanism.
ZK-rollups submit zero-knowledge proofs to the L1 layer attesting to the correctness of transaction execution.
Instead, optimistic rollups, assume L2 blocks to be correct by default, relying on a deferred fraud-proof mechanism to detect invalid states resulting from execution. 
In the general case, when no dispute occurs, the optimistic approach yields lower costs and faster soft confirmation times for transactions (pending finality after the fraud-proof interval) compared to ZK-rollups.
Consequently, optimistic rollups dominate in terms of transacted volume~\cite{l2beat}. 

In most rollups, either optimistic or ZK-based, the \emph{sequencer}~\cite{motepalli2023sok} is a centralized component. 
This maintains a private memory pool (\ie \emph{mempool} for short) of user transactions, ordering, executing, and including them in L2 blocks, which are further compressed in batches submitted to L1. 

\subsection{Proposer-builder separation in L2}

\er{This section gives the impression that we will study decentralized rollups, and that these are a reality. It would make sense to tone it down and explain that PBS is, for the moment, only a solution for modularity in logically centralized sequencers (and thus not a response to the problem of centralized sequencers' deployment). We can say it is a step toward a decentralized PBS setting, but many challenges remain to enable it, and solving them is not the focus of this paper, which explores the current, alternative, logically centralized designs.}
The current prevalent use of a centralized sequencer in rollups undermines the decentralization principle of blockchain systems and allows the sequencer to engage in censorship practices, for economic reasons, e.g., unregulated MEV extraction~\cite{alipanahloo2024maximum}, or political/regulatory reasons~\cite{wahrst24}.  
Besides tampering with legitimate block building, a monolithic sequencer is also a potential single point of failure and performance bottleneck.

Proposer-builder separation (PBS)~\cite{ethereum-pbs} is a design paradigm first adopted in L1 networks~\cite{mev-boost} to mitigate such issues, accounting for the vast majority of blocks on Ethereum ($\approx90\%$~\cite{10.1145/3618257.3624824}). 
In contrast, PBS is not currently implemented in L2 rollups in the same sense as in L1. 
Existing rollup deployments experiment with a limited form of separation in which a single  builder is used de facto alongside the sequencer~\cite{opmainnet-flashblocks, unichain, base-flashblocks}.

This PBS approach in L2 separates block construction, including transaction ordering, from the other operations, such as block compression and batching, rollup state root derivation following the new block, ZK-proof computation for ZK-rollups, and posting these results to L1. 
Depending on the rollup architecture, some of these tasks are either still performed by the sequencer or are delegated to other nodes.
As in L1~\cite{li25}, nonetheless, transaction ordering into blocks is delegated to a builder, as it is highly sensitive to censorship and requires protection~\cite{pbs-tee-opodis25,intro-rollup-boost}.
For this reason, deployed PBS rollup implementations~\cite{opmainnet-flashblocks, unichain, base-flashblocks} adopt the Rollup Boost~\cite{intro-rollup-boost} design, in which the builder operates the mempool, ordering logic, and execution engine within a TEE. 
The foreseen L2 PBS roadmap includes legitimate MEV pipelines similar to L1, and further decentralization to multiple builders~\cite{intro-rollup-boost}.
However, at the time of writing, these implementations adopt only a 1-to-1 separation between the sequencer and the builder, focusing mainly on optimizing builder performance~\cite{op-rbuilder}. 

We set our practical experience report in this logically centralized, albeit modular setting, currently met in production environments. 
Despite the gap towards a truly decentralized context, this builder separation facilitates a first step in enhancing the performance of a censorship-resistant implementation.

\subsection{Trusted execution environments} 

A TEE enables code to be executed in isolation from its operating environment.
It provides \emph{confidentiality} of runtime data, and \emph{integrity}, ensuring code and data cannot be tampered with, not even by privileged software like the OS or hypervisor. 
This is backed by \emph{remote attestation} of the underlying hardware and code integrity. 
In the context of L2 rollups, a TEE can ensure that the underlying algorithms, \eg transaction ordering and block-building logic, adhere to predefined rules. 

TEE technologies generally fall into two categories: \textbf{process-level isolation}, where a portion of a process's memory is protected at runtime (Intel SGX~\cite{vcostan}), or \textbf{virtual machine (VM)-level isolation}, where an entire VM is protected for its lifetime (Intel TDX~\cite{cheng24}, and AMD SEV~\cite{sev-snp}).
VM-based TEEs mitigate the performance and usability issues associated with process-level TEE, but lead to a larger trusted computing base (TCB), \ie the set of software and hardware that must be trusted. 
Current L2 rollups rely on Intel TDX for protection through the Rollup Boost implementation~\cite{intro-rollup-boost}.
All TEEs can introduce variable overheads to protected processes. 

\section{TEE-enabled rollup designs}
\label{sec:arch}

\begin{figure}[!t]
	
	\centering	
	\includegraphics[scale=0.55]{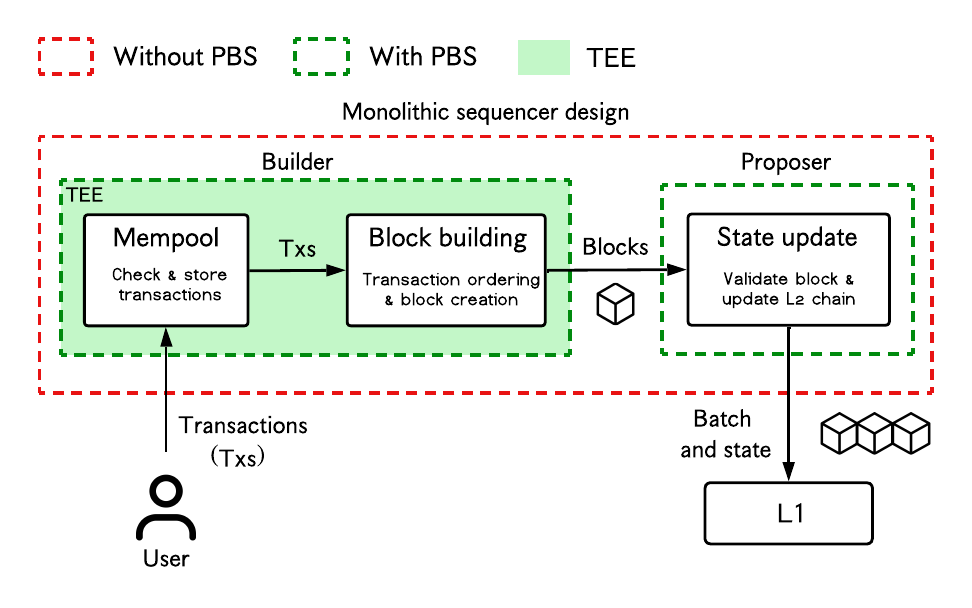}
	\caption{L2 rollup designs with and without PBS. In a PBS design, the builder can be run within a TEE for censorship resistance.}
	\label{fig:pbs-archs}
\end{figure}

Our objective is to identify the performance and security tradeoffs in TEE-enabled rollup designs, whether or not based on PBS.
Independent and systematic measurements of these tradeoffs are lacking in the scientific literature.
We now systematically detail these designs.

We choose to target optimistic rollups, the dominant rollup category by usage and transaction volume~\cite{l2beat}.
The only production-ready PBS approach, currently deployed in L2 systems, is Rollup Boost. 
It integrates with OP Stack, the leading framework for implementing optimistic rollups, making the study of optimistic rollups more immediately relevant to practitioners.
Nonetheless, our study is also of interest to designers of zk-rollups, as it focuses on securing the operational integrity of the private mempool, which is centralized in the majority of optimistic rollups and in many zk-rollups.

We consider two categories of systems: monolithic sequencer designs that do not use PBS, and PBS-based designs that use external block builders for block construction.
\autoref{fig:pbs-archs} provides a high-level description of TEE-enabled rollup designs.

\subsection{Threat model}
We consider a rollup design, where the mempool is private to a node in charge of constructing L2 blocks (\ie a builder or sequencer). 
Each node has control over the full software stack, including the OS, hypervisor, and hardware (\eg, DRAM, secondary storage, \etc).
An adversarial node could aim to engage in (dishonest) censorship actions such as strategically ordering, including, excluding, or frontrunning client transactions that are present in the mempool.
In short, the node privately operating the transaction mempool is capable of and interested in arbitrary changes to the mempool content, deviating from publicly agreed-upon transaction ordering criteria within and across blocks. 
For further analysis of censorship actions, we refer the reader to the formal definitions of Arusoaie \textit{et al.}~\cite{pbs-tee-opodis25}.

\smallskip\noindent\textbf{Assumptions.}
We assume that a TEE-enabled sequencer or builder is pre-attested before sending transactions, using TEE-specific attestation mechanisms. 
We also assume an adversary cannot physically open and manipulate the TEE-enabled processor package, and that the TEE (\eg enclave or VM) code does not intentionally leak sensitive data. 
Denial-of-service and side-channel attacks are considered out of scope.

\subsection{Non-PBS designs}
Here, we consider a monolithic sequencer that may or may not be TEE-enabled. 
In the non-TEE architecture (assumed as the baseline), the sequencer can exclude, reorder, or include transactions in the private mempool, altering the produced blocks.
A TEE-enabled sequencer cannot perform these actions, because clients cryptographically verify the mempool management and ordering code via remote attestation. 

\subsection{Designs with PBS}
In PBS designs, separate nodes handle the block-building and -proposing logic: the block builder and the sequencer (acting as the block proposer).
We consider two cases: one where the builder runs inside a TEE, preventing tampering with mempool operations and transaction ordering~\cite{pbs-tee-opodis25}, and exporting blocks that can be integrity-checked; and one where the builder does not run inside a TEE, thus allowing such tampering.

\section{Evaluation}
\label{sec:eval}

The evaluation seeks to answer the following questions:

\begin{itemize}[leftmargin=33pt,nosep]
	\item[\textbf{RQ-1}:] What is the impact of PBS on metrics such as transaction time to inclusion (TTI), \ie time from submission until inclusion in an L2 block? (\S\ref{eval:tti}) 
	\item[\textbf{RQ-2}:] What is the impact of TEEs on the above metrics with and without a PBS configuration? (\S\ref{eval:tti}) 
    \item[\textbf{RQ-3}:] How does CPU utilization (\S\ref{eval:cpu}) and RAM (\S\ref{eval:ram}) resource consumption vary across the different designs? 
\end{itemize}

\subsection{Evaluation setup}
We emulate a full Optimism rollup network on top of an emulated Ethereum node acting as the L1 chain. 
The setup is based on Flashbots Builder Playground~\cite{builder-playground}, an open-source framework for deploying full blockchain testing environments, that we adapted for Kubernetes compatibility.
Builder Playground provides PBS support via Rollup-Boost~\cite{intro-rollup-boost}, a middleware component that enables the rollup sequencer to outsource block construction to an external block builder.

We use a Kubernetes cluster with 4 VM nodes on a host server. 
The VM host server has a single-socket 16-core Intel(R) Xeon(R) Gold 5515+ CPU clocked at 2.70GHz 
with Intel TDX support, and uses the KVM hypervisor.
The VMs, running Ubuntu 24.04 LTS, have 4 vCPUs and 4GB of memory each, and one of them is TDX-enabled. 
We distribute the work using node affinity rules as follows: one node for metrics collection and load injector, sequencer components on a second node, and the external builder on the third or fourth node, depending on whether the TEE is enabled or not. 

\subsection{Evaluation methodology}
We benchmark the different blockchain designs with Contender~\cite{contender}, a high-performance transaction generator developed by Flashbots.
Contender generates large volumes of transactions from a client VM towards the blockchain's RPC endpoint and measures various metrics, including TTI. 
We evaluate the different designs by repeatedly performing \emph{ERC-20 token transfer}~\cite{erc-20} transactions from a set of test pre-funded accounts to a random recipient address. 
ERC-20 transactions are among the most common in Ethereum blockchains~\cite{somin2025cryptoasset}.
Additionally, we also observe the median results for simulated loads of~\emph{Uniswap} transactions. 
Uniswap is currently the most-deployed decentralized exchange protocol, enabling trading of ERC-20 tokens, with a presence on 44 chains and 747 forks for Uniswap V2~\cite{defillama-uniswap}.
We provide results for both low (30, 90) and high (210, 270) input transaction-per-second (TPS) levels.  
We note that the daily average volume in real-world layer-2 rollups such as Optimism is in the range of 20 to 30 TPS. 

With monolithic sequencer configurations, we use Optimism's reference builder implementation for the L2 stack (\opgethbuilder)~\cite{op-geth}. 
For the PBS setups, we use \opgethbuilder as an external builder connected to the L2 stack through Rollup-Boost.
We also evaluate our PBS designs with \oprbuilder~\cite{op-rbuilder}, Flashbots' block builder implementation for the L2 stack.
This allows us to evaluate PBS performance across different builder implementations. 
All our PBS designs use a single external builder, which, to our knowledge, is the common practice in deployments using Rollup Boost as of the date of writing.

In our evaluations, we adopt the following naming: 
\begin{enumerate}
    \item Monolithic sequencer without a TEE: \nopbsnotee 
    \item Monolithic sequencer running in a TEE: \nopbstee 
    \item PBS with a non-TEE enabled builder: \pbsnotee 
    \item PBS with a TEE-enabled builder: \pbstee 
\end{enumerate}

PBS systems are further compared with respect to the integrated builder implementation: the suffix ``\texttt{-rb}'' indicates that \oprbuilder is the external builder used for the corresponding PBS setting (\eg ``\pbsteerb'').
Otherwise, the builder implementation is \opgethbuilder (same as used in the monolithic sequencer setting).
In addition to L2 chain metrics, we measure system-level metrics such as CPU usage and RAM usage of the block builders in all setups, collected with Prometheus. 

\subsection{Impact of PBS and TEEs on L2 performance}
\label{eval:tti}

\begin{figure*}[!t]
	\centering
	\includegraphics[scale=1.10]{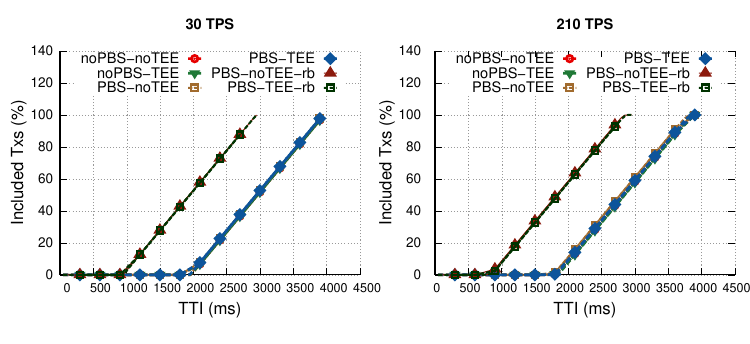}
	\caption{Cumulative distribution function (CDF) of Tx TTI for different TPS values using ERC-20 workload. We note the almost-overlapping results for each of the two builders, \opgethbuilder (the first four lines in the legend) and \oprbuilder (the last two lines with the -rb suffix).}
	\label{fig:eval-tti}	
\end{figure*}

We start by observing in \autoref{fig:eval-tti} the distribution of TTIs (\ie time to inclusion). 
The plots depict the results obtained with an ERC-20 workload at 30 and 210 TPS, running for 10 minutes.
The Uniswap workload provided highly similar results. 
The main difference in our tests was inflicted by the type of builder used. 
We observe in all runs that use \opgethbuilder, that the time-to-inclusion of transactions is near-identical for both the non-PBS and PBS variants, regardless of whether TEEs are used or not.
Similarly, we do not observe a significant effect of TEE in the tests run with \oprbuilder.

The very small TTI overhead introduced by VM-based TEEs reflects the results in~\cite{eval-tee-types}, which evaluates both TDX and SGX. 
That work implies that the impact of TEEs on operational performance is generally small for CPU-intensive workloads. \emi{we can't get something obvious on this from the current plots; I've added a "very" before small overhead, but maybe we should provide some numerical value, \eg some average of the difference}
In contrast, highly I/O-intensive operations tend to incur higher overhead due to TEE-related context switches. 
Block building in the studied L2 chain setup is predominantly CPU-intensive, with comparatively limited I/O. 
This explains the minimal performance impact of Intel TDX on the block-building operation and, hence, on TTI, in our experiments.

\observ{When using the same builder implementation for non-PBS and PBS designs, VM-based TEEs have a negligible impact on transaction time to inclusion. This is mainly because block building is CPU-intensive, incurring minimal overhead when run in a VM-based TEE.}

Regarding PBS, given that the external builder runs on a VM on the same host, network latency's impact is small.
In a real-world PBS setting, the external builder will typically be on a separate node, leading to higher network latency. 
The latter is expected to increase the TTI of transactions in blocks. 

However, we also observe that when \oprbuilder is used as the external builder in PBS settings, the TTI is twice as fast as in PBS setups using \opgethbuilder.
We attribute these performance differences primarily to the architectural and implementation differences between the two builder implementations \opgethbuilder and \oprbuilder.
On the one hand, \oprbuilder is a lightweight and more specialized Rust-based block builder comprising $\approx14$K lines of code (LoC).
On the other hand, \opgethbuilder is a significantly more complex Go-based system comprising $\approx 352$K LoC. 
In addition to block building, \opgethbuilder also participates in block validation.

\observ{
PBS enables the use of more specialized and efficient block builder implementations, which can lead to significant performance improvements when compared to a monolithic sequencer setting.
}

\subsection{Impact of PBS and TEEs on CPU utilization}
\label{eval:cpu}

\begin{figure*}[!t]
	\centering	
	\includegraphics[scale=1.1]{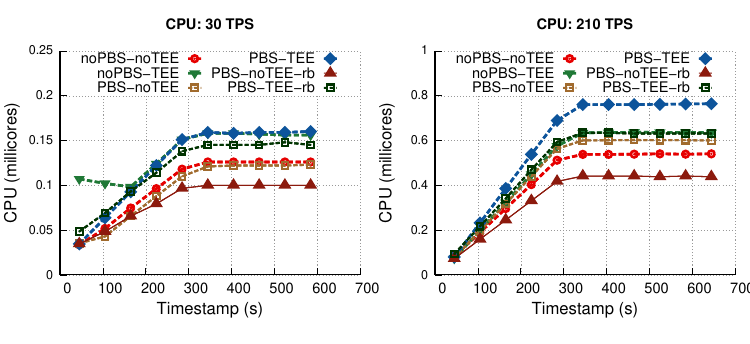}
    \includegraphics[scale=1.1]{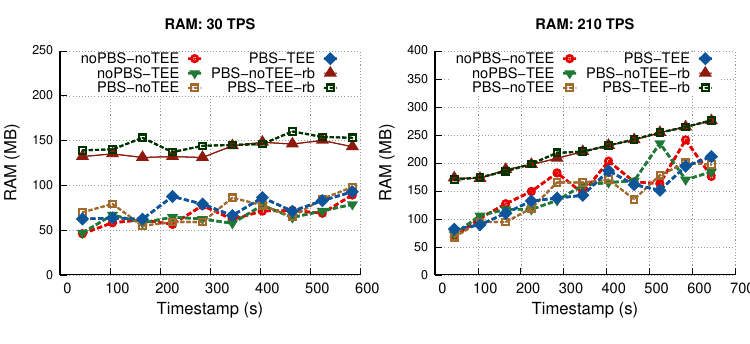}
	\caption{CPU and RAM usage for block builder across different setups using ERC-20 workload (1 CPU = 1000 millicores). The suffix "\texttt{-rb}" indicates that the builder used is \oprbuilder; otherwise, the default builder, \opgethbuilder is used.}
	\label{fig:eval-cpu}	
\end{figure*}

We now look at the impact of the different designs on CPU utilization.
The top plots in \autoref{fig:eval-cpu} depict the CPU usage for the components involved in block building in both PBS and non-PBS settings, with and without TEEs, under ERC-20 load.
The results for the Uniswap workload are presented in \autoref{fig:eval-ram}.

We observe that CPU usage increases in TEE-based setups compared to non-TEE settings. 
For example, with a load of 210 TPS, \nopbstee uses about $16.7\%$ more CPU compared to \nopbsnotee, while \pbstee uses approximately $25.3\%$ more CPU compared to \pbsnotee.
We observe a similar trend for the 270 TPS Uniswap workload, albeit more stable in the first part of the interval.
We believe this variation stems from differences in the processing logic of the two types of transactions in the load. 
We attribute the higher CPU utilization of the TEE-based designs to underlying hardware-based cryptography.
In VM-based TEEs such as Intel TDX, cryptographic operations, \eg memory encryption and decryption of pages between the CPU and main memory, incur additional processing cost. 
Even though block building is primarily CPU-bound, some data is still copied between memory and the processor, which requires additional CPU cycles for cryptographic operations.
Consequently, the TEE-based configurations exhibit moderately higher CPU utilization.
Nonetheless, this CPU usage overhead has a negligible impact on TTI, as previously illustrated in \autoref{fig:eval-tti}.

\observ{
TEEs increase CPU usage by about $17\%$ and $25\%$ for non-PBS and PBS setups, respectively. This can be attributed to the additional CPU cycles required to perform hardware-based cryptographic operations in TEEs.
}

From \autoref{fig:eval-cpu}, we also observe that under the higher ERC-20 load of 210 TPS, PBS-based setups exhibit higher CPU utilization than their non-PBS counterparts.
For instance, \pbsnotee uses $10.9\%$ more CPU than \nopbsnotee. 
This CPU usage trend is also observed with the Uniswap workload in \autoref{fig:eval-ram}, stabilising after about 3 minutes of the tests duration.
This increase can be attributed to the additional network communication introduced by PBS, in which transactions must be forwarded to an external builder.
Handling these extra network interactions adds computational work for the node, thereby increasing overall CPU usage. 
This overhead becomes dominant once the processing stabilises with the incoming transactions, reaching a plateau.
In a brief initial part of the tests interval, it can be observed that the non-PBS CPU utilization is generally higher due to the concentration of processing logic in the monolithical setting.

\observ{
In PBS settings, the required network communication with an external builder increases CPU utilization by the components responsible for block construction.
}

\subsection{Impact of PBS and TEEs on RAM utilization}
\label{eval:ram}

\begin{figure*}[!t]
	\centering	
    \includegraphics[scale=1.1]{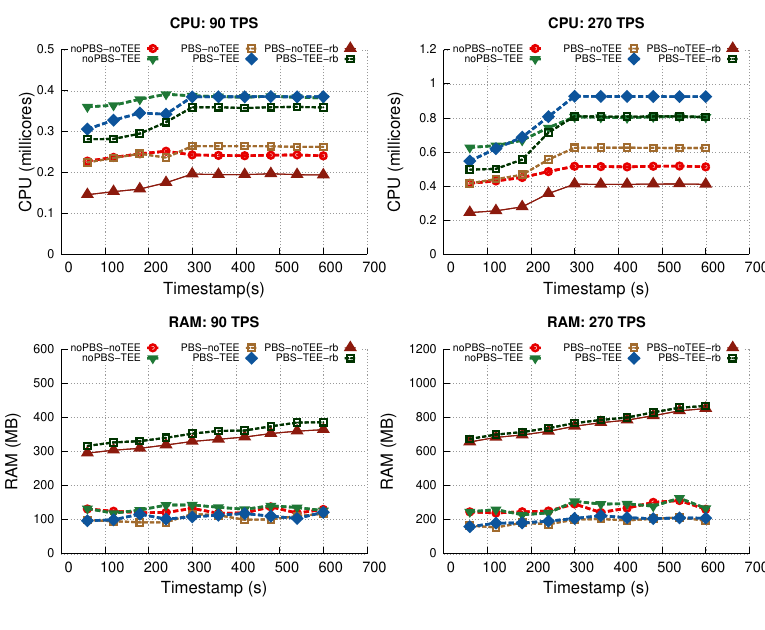}
	\caption{CPU and RAM usage of the block builder across different setups using Uniswap workload. The suffix "\texttt{-rb}" indicates that the builder used is \oprbuilder; otherwise, the default builder, \opgethbuilder is used.\vs{for TEE configs, what is 'memory'?}\py{I don't understand your question.}}
	\label{fig:eval-ram}	
\end{figure*}

Finally, we look at the impact on memory usage.
The bottom plots in \autoref{fig:eval-cpu} and \autoref{fig:eval-ram} show RAM usage for the components involved in block building for lower TPS ERC-20 loads and higher TPS Uniswap loads, respectively.

In general, for the same builder implementation, we observe no clear variation in RAM usage between PBS and non-PBS settings, regardless of TEE usage. 
This can be explained by the fact that all evaluated designs maintain similar in-memory data structures regardless of whether PBS or TEEs are used, resulting in comparable RAM usage in the guest VM.
However, we note that TEEs such as Intel TDX store additional cryptographic metadata in a separate memory region rather than the guest VM's memory. 

Notably, for input loads of 30 TPS and 210 TPS, the RAM usage of \oprbuilder is, respectively, about $1.927\times$ and $1.597\times$ higher on average than that of \opgethbuilder in the PBS setting. 
Again, we observe a similar tendency for the Uniswap workloads at 90 TPS and 270 TPS. 
We attribute this difference primarily to implementation-level design differences rather than the programming language itself (\ie Rust-based programs, \eg \oprbuilder, generally use less memory than Go-based ones, \eg \opgethbuilder). 
In particular, the two builders rely on different internal mechanisms to manage the transaction pool, cache execution state, and maintain intermediate EVM data structures during block construction. 
These differences can affect how aggressively state or transaction data are cached in memory, resulting in higher RAM usage for \oprbuilder in our experiments. 
A detailed breakdown would require fine-grained memory profiling of the respective implementations, which we leave for future investigation.
Nonetheless, the RAM usage is the same across all setups using \oprbuilder \ie, \textbf{\texttt{PBS-noTEE-rb}} and \textbf{\texttt{PBS-TEE-rb}}, which is consistent with the results with \opgethbuilder and equally explained by the fact that the in-memory data structures and how much memory they occupy in the guest VM are the same when the input load and block builder implementation remain constant. 
We also note that TDX uses AES-XTS~\cite{aes-xts} as a symmetric block cipher for memory encryption, where the ciphertext does not expand on the plaintext size, which is in line with the observed results. 

\observ{
RAM utilization is largely independent of PBS and TEE usage and instead dominated by the block builder implementation. While TDX introduces additional cryptographic metadata, it resides outside the guest VM memory and does not increase the observed RAM footprint.}

Lastly, similar to the CPU usage patterns under higher loads, we also observe higher RAM usage across all setups at the higher input loads of 210 and 270 TPS than at the lower input loads of 30 and 90 TPS.
The higher loads increase RAM usage because more transaction data is processed per unit time compared to the lower loads.
We also note a gradual increase in overall RAM consumption over time for each executed test, more noticeable in the higher TPS runs.
We believe this is mainly caused by the increase in blockchain size: new transactions result in new blocks and changes to the blockchain state, which are typically kept, at least temporarily, in memory by implementations such as \opgethbuilder and \oprbuilder to optimize new block additions and transaction execution.
\emi{added the two phrases above - I think there was some official info about Geth maintaining the last 128 blocks cached in memory + intermediate states for fast operation in case of reorgs, but I didn't recheck that, and anyway implementation might differ in OP Stack}

\subsection{TEE Integration Discussion}

We observe that a baseline TDX integration with a block builder does not cause a major overall performance degradation, favoring the use of the TEE in rollup solutions. 
Therefore, we discuss TEE integration aspects beyond this baseline.

\smallskip\noindent\textit{Choice of TEE.}
VM-based TEEs like Intel TDX and AMD SEV-SNP are the growing trend in confidential computing due to better usability and improved performance compared to process-based TEEs like Intel SGX.
The complexity of builder implementations, \eg \oprbuilder~\cite{op-rbuilder} ($\approx14$K lines of code (LoC)), \opgethbuilder ($\approx 352$K LoC), suggests VM-based TEEs as the appropriate solution for TEE-assisted security in this domain, as opposed to the process-based TEE solutions which require extensive code refactoring and which impose large performance hits.
Nonetheless, reducing the larger TCB of VM-based TEE designs remains an open research problem.

\smallskip\noindent\textit{TCB reduction.}
TCB reduction in VM-based TEEs remains a fundamental requirement for improving security. 
One promising direction is the use of unikernel-based designs, where the VM image is specialized for a single application or service, eliminating unnecessary OS components and thus reducing the attack surface.
Recent work~\cite{incognitos} explored this direction.
A complementary approach is to use a LibOS architecture within VM-based TEEs, such as Gramine-TDX~\cite{gramine-tdx}, which could also provide specialized VM images for the block builder.

\smallskip\noindent\textit{Code attestation with VM-based TEEs.}
In VM-based TEE designs like Intel TDX, attestation mechanisms primarily focus on boot-time components such as the kernel and firmware. 
As a result, boot-time attestation alone may not capture the integrity of application-level software such as the builder or sequencer, unlike process-based TEEs like Intel SGX. 
To address this limitation in VM-based TEE attestation, technologies such as the Integrity Measurement Architecture (IMA)~\cite{linux-ima}, introduced in the Linux kernel, can be leveraged.
IMA records measurements of executed binaries and maintains an integrity measurement log in TEE-protected memory, enabling verification of runtime components.
Since measurements can be limited to the first execution of binaries via IMA policy rules, the runtime overhead introduced by IMA for components such as the builder is not significant.

\section{Related work}
\label{sec:rw}

\smallskip\noindent\textbf{PBS.}
Most prior work on PBS focuses on L1 blockchains.
Heimbach \textit{et al.}~\cite{10.1145/3618257.3624824} analyze PBS in L1 across several dimensions, including decentralization in transaction validation and censorship resistance.
Yang \textit{et al.}~\cite{yang2025} investigate L1 PBS, highlighting the implications of MEV centralization in block builders.
Our work tackles PBS in L2, evaluating the performance impact of both TEEs and PBS in rollup designs.

\smallskip\noindent\textbf{TEE-based censorship solutions.}
To our knowledge, Flashbots' Rollup Boost~\cite{intro-rollup-boost} is the only deployed solution that combines TEEs with PBS in the context of L2, and no prior work has provided a thorough experimental evaluation of this PBS design.
This is currently integrated with OP Stack, the framework used in the OP Mainnet rollup~\cite{opmainnet-flashblocks}, and also adopted in several other rollups~\cite{unichain, base-flashblocks}.
Some past studies have explored TEE in the context of censorship in L2: TEE Rollup~\cite{DBLP:journals/tc/WenFLNZF25} proposes a network of TEE-enabled monolithic sequencers to address both censorship and fault tolerance at the Rollup sequencer level. 
The empirical evaluation of their system shows that TEEs decrease transaction throughput by $15.74\%$ and increase transaction latency by $20.17\%$. 
However, this evaluation is performed in the context of multiple heterogeneous sequencers, both VM-based and process-based, that also run a quorum-based coordination process to choose the block proposer.
Cristiano \textit{et al.}~\cite{Cristiano2025EnhancingTS} propose a TEE-enabled sequencer design equipped with a decentralized attestation mechanism.
Their performance evaluation with Intel SGX shows that Intel SGX increases transaction confirmation time by $\approx 614\% - 740\%$. 
Li \textit{et al.}~\cite{li25} propose TEE-based mempools as a means to achieve transaction fairness in L1 blockchains, using again only Intel SGX.


\section{Conclusion}
\label{sec:conclusion}
In this paper, we empirically evaluate TEE-enabled rollup designs under both PBS and non-PBS configurations.
Our work provides insights into the separate impact of TEEs and PBS, not previously evaluated comparatively, in the context of rollup designs currently deployed in practice, i.e., based on Rollup Boost. 
Our results show that PBS introduces some network communication overhead, whose performance impact can be further amplified by TEEs, albeit moderately, and is mainly limited by CPU usage. 
Nonetheless, the TEE-integrated PBS case is the main step towards a safe, censorship-resilient rollup.
In this architecture, our evaluations show that PBS can improve performance by enabling more specialized and efficient builder implementations.
PBS can also mitigate any performance impact of TEEs on other complex operations, such as batching and state computation at the sequencer, which runs outside a TEE in PBS designs.
We consider finer-grained benchmarking, covering the externalization of other sequencer functionalities, as future work.

\bibliographystyle{IEEEtran}
\bibliography{references} 

\end{document}